\documentclass{udine} 
\usepackage{epsfig,epsf}

\begin{document}

\title{GLAST, the Gamma-ray Large Area Space Telescope}

\author{Alessandro de Angelis}

\address{Universit\`a di Udine and INFN Trieste, via delle Scienze 208, I-33100 Udine (Italy)\footnote{also at Instituto Superior T\'ecnico, Lisboa.}\\E-mail: deangelis@ud.infn.it}

\maketitle

\abstracts{GLAST, a detector for cosmic $\gamma$ rays
in the range from 20 MeV to 300 GeV, will be launched in space in 2005.
 Breakthroughs are expected in particular in the
study of particle acceleration mechanisms in space and 
of gamma ray bursts, and maybe on the search for cold dark matter; but
of course the most exciting discoveries could come from the unexpected.}

\section{Introduction}

The exchange between laboratory physics and astrophysics has demonstrated
through centuries of science to be at the origin of fundamental discoveries.
GLAST\cite{GLAST}, a detector for cosmic $\gamma$ rays
in the range from 20 MeV to 300 GeV, is built around this idea: both the
design of the
instrument and the guidelines for science originate from a partnership of
researchers from High Energy Physics (HEP) and High Energy Astrophysics (HEA).
The collaboration involves the US (NASA in particular), 
France, Italy, Japan and Sweden.

The study of $\gamma$ rays is fundamental for our understanding 
of the universe\cite{reviews}: $\gamma$ rays 
probe the most energetic phenomena 
occurring in nature, and several signatures of new physics are associated 
with the emission of $\gamma$ rays. 
Photons can travel essentially undeflected and unabsorbed in space,
and thus they point with excellent approximation to the source of their
emission.
Moreover, we have the technology for building high-performance
$\gamma$ detectors.
The design of GLAST, which will be better described in the next section, is 
rather typical for a modern HEP detector: essentially a silicon tracker 
followed by a calorimeter.  

GLAST comes after the scientific success of the Energetic Gamma Ray Experiment
Telescope (EGRET) instrument on the Compton
Gamma Ray Observatory\cite{EGRET}. Launched in 1991, EGRET
made the first complete survey of the sky above
30 MeV. EGRET increased the number of identified $\gamma$ sources
producing a catalog which is a reference. Although EGRET answered many
questions, it opened as many mysteries, which GLAST hopes to solve.
GLAST maintains the same design philosophy as EGRET, but 
will improve substantially the 
EGRET performance by using up-to-date technology from HEP.

GLAST will be sent in space in 2005; it should orbit at a height
of 550 Km and make a full sky coverage in 2-3 orbits,
sending data at a rate of 0.3 Mb/s (which could triplicate if the
rapidly growing telecommunication
technology will allow it). The detector is expected to take data for at least 
five years, with the possibility of extending the mission to ten years.
The launch by a Delta 2 vector limits the weight of GLAST to around 4 tons 
and its size to 2x2x1 m$^3$. The total power consumption is limited to
650 W.

\section{The detector}

\begin{figure}
\center{\epsfig{file=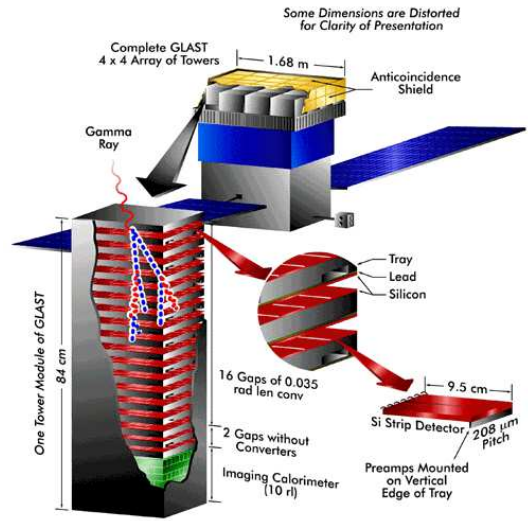,width=0.6\linewidth}}
\caption{GLAST in detail.}\label{fig2}
\end{figure}

GLAST is an array of 4x4 identical towers (Figure \ref{fig2}), embedded in
a plastic scintillator shield which serves as an anticoincidence
trigger.
Each  tower, about 40x40 cm$^2$ in surface, comprises a tracker, 
a calorimeter and a data acquisition module.

The tracking detector consists of 18 $xy$ planes of silicon strip 
detectors, with pitch of about 200 $\mu$m. On 16 of them a converter will
increase the radiation length so to increase the probability of $\gamma$ 
conversion; the electron-positron pair coming from such conversion can be
tracked to reconstruct with accuracy the direction
of the incoming photon. 
The choice of silicon guarantees
high signal/noise ratio, radiation hardness and
low power consumption. The system is mechanically stiffened by 
carbon walls; electronics is placed on the sides
to minimize the gaps. 

The calorimeter in each tower consists of 12 CsI bars in a hodoscopic
arrangement, read out by double photodiodes; the total thickness is 10
radiation lengths. The hodoscopic
geometry, with alternate $x$ and $y$ planes,  
allows fitting the shower profile, so to
be able to compute the leakage and overcome the consequent
deterioration of the energy resolution.

\begin{figure}
\center{\epsfig{file=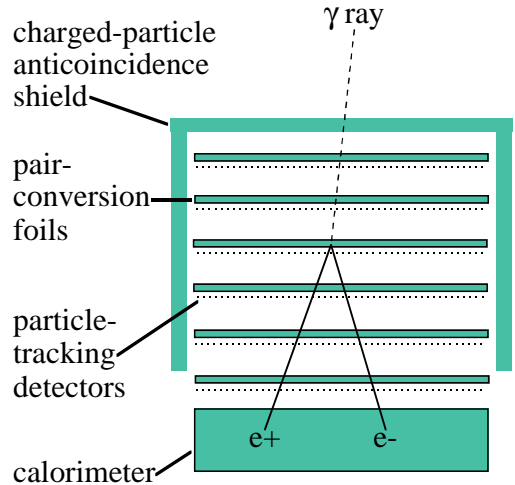,width=0.6\linewidth}}
\caption{Principle of photon detection in GLAST.}\label{fig3}
\end{figure}

The setup consisting of a tracker and a calorimeter is a standard for the 
modern HEP experiments: the tracker detects the conversion and 
allows fitting
the photon
direction, while the calorimeter measures the total energy
(Figure \ref{fig3}).
The anticoincidence detector helps in suppressing the background from
charged cosmic rays, which is five orders of magnitude larger than the 
$\gamma$ signal. 
The height/width ratio of $\approx 0.4$ guarantees a large 
field of view; 
the arrangement in 16 complete towers guarantees modularity.

As a result GLAST, built around the same design philosophy as EGRET,
can increase substantially EGRET's sensitivity (Table \ref{table1}).
\begin{table}
\begin{center}
\begin{tabular}{lll}\hline
 & {EGRET} & {GLAST} \\ \hline
Energy range &	0.02--30 GeV &	0.02--300 GeV \\
Energy resolution $\Delta E/E$  & 10\% ($E>100$ MeV) & 10\% ($E>100$ MeV)  \\
Single $\gamma$ angular resolution & 5.8$^\circ$ (100 MeV) & $\simeq3^\circ$
(100 MeV)\\
 & & $\simeq 0.15^\circ$ (10 GeV)\\
Peak eff. area	& 0.15 m$^{2}$ &	1 m$^{2}$ \\
Field of view &	0.5 sr &	2.4 sr \\
Time resolution & 100 $\mu$s & 10 $\mu$s\\ \hline
\end{tabular}
\caption{Selected performance parameters for GLAST and EGRET.}
\label{table1}
\end{center}
\end{table}

\begin{figure}
\center{\epsfig{file=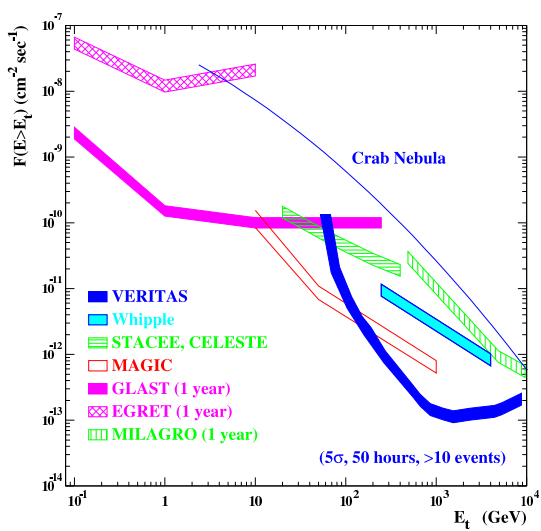,width=0.6\linewidth}}
\caption{Sensitivity of present and future 
high energy $\gamma$ detectors.}\label{sensgamma}
\end{figure}

GLAST is complementary to ground-based experiments, and is a key of the 
$\gamma$ astrophysics program
for the new century (Figure \ref{sensgamma}).

\section{Key science objectives}

GLAST has a wide range of physics objectives, ranging from
gamma astrophysics to fundamental physics. For a more complete review,
see\cite{GLAST} and references therein. I try here just to give a
short summary. 

\begin{itemize}

\item {\em Resolving the $\gamma$ ray sky: 
AGN, diffuse emission and unidentified sources}

EGRET has discovered $\simeq 80$ Active Galactic Nuclei
(AGN) at $\gamma$ ray energies, 
and GLAST will discover several thousands; the 
$\gamma$ part of the spectrum could be dominant,
and the variability in time of the sources will provide important information.

Most of the gamma-ray sources
detected by EGRET remain unidentified; the accuracy of EGRET could allow
identifying such sources.

Besides the AGNs, EGRET discovered a diffuse background radiation.
Is it really diffuse, and thus  
produced at a very early epoch, or a flux from unresolved sources? 
GLAST will improve the angular resolution and the sensitivity to weak sources,
possibly giving an answer.

\item {\em Acceleration mechanisms of Cosmic Rays (CR)}\\
The signal of $\pi^0$
decaying into $\gamma\gamma$ could indicate the
dominant role of the acceleration of nuclei.
The angular resolution of GLAST will allow
to possibly associate CR sources to supernova remnants.

\item{\em Gamma-Ray Bursts (GRB)}\\
EGRET detected high energy afterglows (which can last for more than 
one hour) in 2-3 out of 6 GRBs observed; 
GLAST can provide measurements over a new energy range
for 50-100 GRB/year. The notification of GRBs 
can be given by an on-board trigger (Gamma-Ray Burst Monitor)
in a few seconds, and from earth in a few minutes.

In addition, recent quantum gravity models predict that the speed of photons
depends on their energy\cite{qg}; this effect could cause delays
of O(100ms) in the arrival time of photons from GRBs, and thus be detectable.

\item {\em Solar flares}\\
GLAST could image these unexplained solar explosions, measuring the
surface interested.

\item {\em Probing dark matter: WIMPs}\\
If dark matter is made by Majorana particles weakly interacting
and with thermic energies,  
the annihilation of such particles
should result in narrow $\gamma$ lines.
This is in particular true in SUSY, where the Lightest Supersymmetric
Particle (LSP) can annihilate with itself producing $\gamma\gamma$ or
Z$\gamma$ pairs. GLAST can be sensitive to a non-negligible part of the
space of the Minimal Supersymmetric Standard Model (MSSM) parameters
for a LSP in the range from 30 to 100 GeV/$c$.

\item{\em Last but not least, the totally unexpected}\\ 
could come from the newly opened exploration regimes.

\end{itemize}

\section{Status of GLAST}

The GLAST design has been extensively studied by means of simulations, and 
test beams are well on the way since three years to verify the 
expected performance.

In particular, the possibility of the calorimeter to reach a relative
energy resolution $\Delta E/E \simeq$ 10\% by fitting the shower profile 
has been successfully verified by dedicated tests in 1997. 

A full GLAST tower has been tested at SLAC in 1999/2000 with incident electron, photon and hadron beams.
It has been verified in
particular that a rejection factor of $10^5$ against hadrons is realistic, 
still
keeping 80\% of the photons.

The software for the simulation, the analysis and the distribution
of data is being developed along the line of 
modern data handling technologies;
simulation and reconstruction are well advanced and being used on 
the test data. A particular mention to the simulation: 
the detector project has been done with detailed simulations
based on GISMO\cite{GISMO}, and now a simulation is being
developed between Europe and Japan in close collaboration with the 
CERN Geant4\cite{G4} development team,
which profits also of the interaction with
ESA.

\section{Summary and conclusions}

GLAST will provide an important 
step forward in $\gamma$ astronomy,
increasing the sensitivity of EGRET by one-two orders of magnitude.

The progress is in schedule for the launch in 2005; beam tests
and software developement are well on the way.

We are confident that this partnership between HEP and 
HEA will be the source of new discoveries over a wide range 
of subjects.

\end{document}